\documentclass[preprint,showpacs,showkeys]{revtex4-1}
\usepackage{graphicx}
\usepackage[caption=false]{subfig}
\usepackage{graphicx,amsmath}
\usepackage{epsfig}
\usepackage{amssymb}
\usepackage{amsmath}
\usepackage{lipsum}
\usepackage{subfig}
\usepackage{mathrsfs}
\usepackage{booktabs}
\usepackage{tabularx}
\RequirePackage[colorlinks,citecolor=blue,urlcolor=magenta,linkcolor=magenta]{hyperref}
\begin{document} 
\title{Froissart Bound and Transverse Momentum Dependent Parton Distribution Functions (TMDPDF) with self-similarity}
\author{Baishali Saikia$^1$}
\email[Corresponding author : \ ]{baishalipiks@gmail.com}
\author{D K Choudhury$^{1,2}$}
\affiliation{$^1$Department of Physics, Gauhati University, Guwahati- 781 014, Assam, India}
\affiliation{$^2$Physics Academy of North-East, Guwahati 781 014, Assam, India}

\begin{abstract}
Froissart Bound implies that the total proton-proton cross-section (or equivalently structure function) cannot rise faster than the logarithmic growth $\log^2 s \sim \log^2 (1/x)$, where \textit{s} is the square of the center of mass energy and \textit{x} is the Bjorken variable. In the present report, compatibility of such behavior will be shown in case of Transverse Momentum Dependent Parton Distributions Functions (TMDPDF) or TMDs for the model of proton based on self-similarity.\\\\
Keywords: Froissart bound, self-similarity, TMDPDF
\end{abstract}
\pacs{12.38.-t, 05.45.Df, 13.60.Hb, 24.85.+p}

\maketitle

\section{Introduction}
One of the cornerstones of the present strong interaction physics is the Froissart theorem \cite{fe}. It declares that the total cross section of any two-hadron scattering cannot grow with energy faster than $(\log s)^2$ where \textit{s} is the center of mass energy square. Later it was improved by Martin \cite{mrt2,mrtn,yjp}. The original derivation of Froissart \cite{fe} is based on Mandelstam representation and that of Martin \cite{mrt2,luu} is on axiomatic field theory which could be considered as more general. The approach has led further development of the subject \cite{roy,royk,mrt3, roy1, roy3} as well as construction of several phenomenological models \cite{phm1,phm2}. It is therefore as familiar as Froissart-Martin bound. 

Precession measurement of proton-proton (\textit{pp}) cross-section at LHC \cite{lhc1,lhc2,lhc3,lhc4} and in cosmic rays \cite{cosmi} have led the PDG group \cite{pdg} to fit the data with such $\log^2 s$ term together with an additional constant $\sigma \sim A +B \log ^2 s$. There is  also an alternative fit for pp data \cite{roh1} with an addition of non leading $\log s$ term

Exact proof of Froissart Saturation in QCD is not yet been reported. However, in specific models, such behavior is found to be realizable. Specifically, soft gluon resummation models in the infrared limit of QCD \cite{roh2} and /or gluon-gluon recombination as in GLR \cite{glrr} equation or color glass condensate \cite{roh3, roh4, roh5} models  such $\log^2 s$ rise of proton proton cross section is achievable.

In DIS, when Froissart bound is related to the nucleon structure function $F_2(x,Q^2)$, it implies a growth limited to $\log^2\dfrac{1}{x}$.

It is well known that the conventional equations of QCD, like DGLAP \cite{o2,o3,o1} and BFKL approaches \cite{o6,o4,o5,o7}, this limit is violated; while in the DGLAP approach, the small-\textit{x} gluons grow faster than any power of $\ln \left( \dfrac{1}{x}\right)  \approx \ln \left(\dfrac{s}{Q^2} \right)$ \cite{rg}, in the BFKL approach it grows as a power of $\left( \dfrac{1}{x}\right) $ \cite{o6,o4,o5,o7,o8}.

However, in recent years, the validity of Froissart Bound for the structure function at phenomenological level has attracted considerable attention in the study of DIS, mostly due to the efforts of Block and his collaborators \cite{bff,blooo,blo,buu,bu}. 

It was argued in Ref. \cite{buu} that as the structure function $F_2^{\gamma p}(x,Q^2)$ is essentially the total cross section for the scattering of an off-shell gauge boson $\gamma^*$ on the proton, a strong interaction process up to the initial and final gauge boson-quark couplings and Froissart bound makes sense. On this basis, one analytical expression in \textit{x} and $Q^2$ for the DIS structure function has been suggested in \cite{blooo} which has expected Froissart compatible $\log^2 \dfrac{1}{x}$ behavior and valid within the range of $Q^2$: $0.85\leq Q^2 \leq 1200$ GeV$^2$ of the HERA data. Using this expression as input at $Q_0^2 = 4.5$ GeV$^2$ to DGLAP evolution equation, the validity is increased upto 3000 GeV$^2$ \cite{blo}. The approach has been more recently applied in the Ultra High Energy (UHE) neutrino interaction, valid upto ultra small \textit{x} $\sim 10^{-14}$ \cite{bu} . It is therefore of interest to study if such Froissart saturation like behavior can be incorporated in any other proton structure functions as well and can be tested with data.

We studied the possibility of incorporating Froissart saturation like behavior in the parametrization of structure function of nucleon based on self-similarity \cite{dkc} as suggested by Lastovicka \cite{Last} and more recently in improved version Ref \cite{bsd}.

The aim of the present paper is to study the incorporation of Froissart bound in case of Transverse Momentum Dependent Parton Distribution Functions (TMDs) in the models of proton based on self-similarity reported in Ref \cite{bsd,bs1}.

In section \ref{forma}, we discuss the formalism of the work. Section \ref{res} contain the results while section \ref{conc} contains the summary.

\section{Formalism}
\label{forma}
The method of construction of self-similarity based models of Parton Distribution Functions (PDFs) has already been discussed in Ref \cite{bsd,bs1,bsc}. We will outline it for completeness.

\subsection{Models of PDFs based on self-similarity}
\subsection*{Model 1}
The self-similarity based models of the proton structure function suggested by Lastovicka Ref\cite{Last} is based on parton distribution function(PDF) $q_i(x,Q^2)$. Choosing the magnification factors $M_1= \left(1+\dfrac{Q^2}{Q_0^2}\right)$ and $M_2= \left(\dfrac{1}{x}\right)$, the unintegrated Parton Density (uPDF) can be written as \cite{Last,DK6} 
\begin{equation}
\label{E1}
\log[M^2.f_i(x,Q^2)]= D_1.\log\dfrac{1}{x}.\log\left(1+\dfrac{Q^2}{Q_0^2}\right)+D_2.\log\dfrac{1}{x}+D_3.\log\left(1+\dfrac{Q^2}{Q_0^2}\right)+D_0^i
\end{equation}
\\
where \textit{x} is the Bjorken variable and $Q^2$ is the renormalization scale and \textit{i} denotes a quark flavor. Here $D_1,\ D_2,\ D_3$ are the three flavor independent model parameters while $D_0^i$ is the only flavor dependent normalization constant. $M^2$ is introduced to make (PDF) $q_i(x,Q^2)$ as defined below (in Eq \ref{E2}) dimensionless which is set to be  as 1 GeV$^2$ \cite{DK6}. 
The integrated quark densities (PDF) $q_i(x,Q^2)$ then can be defined as
\begin{equation}
\label{E2}
q_i(x,Q^2) = \int_0^{Q^2}f_i(x,Q^2)dQ^2
\end{equation}
\\
As a result, the following analytical parametrization of a quark density is obtained by using Eq(\ref{E2}) \cite{DK5}\\
\begin{equation}
\label{E3}
q_i(x,Q^2) = e^{D_0^i}f(x,Q^2)
\end{equation}
where
\begin{equation}
\label{E4}
f(x,Q^2)= \frac{Q_0^2 \ \left( \frac{1}{x}\right) ^{D_2}}{M^2\left(1+D_3+D_1\log\left(\frac{1}{x}\right)\right)} \left(\left(\frac{1}{x}\right)^{D_1\log \left(1+\frac{Q^2}{Q_0^2}\right)} \left(1+\frac{Q^2}{Q_0^2}\right)^{D_3+1}-1 \right)
\end{equation}
\\
is flavor independent. Using Eq(\ref{E3}) in the usual definition of the structure function $F_2(x,Q^2)$, one can get
\begin{equation}
\label{E5}
F_2(x,Q^2)=x\sum_i e_i^2 \left( q_i(x,Q^2)+ \bar{q}_i(x,Q^2)\right) 
\end{equation}
or it can be written as
\begin{equation}
\label{E6}
F_2(x,Q^2)=e^{D_0}xf(x,Q^2)
\end{equation}
\\
where 
\begin{equation}
\label{Ea}
e^{{D_0}}=\sum_{i=1}^{n_f}e^{2}_{i}\left(e^{D_0^i}+ e^{\bar{D}_0^i}\right)
\end{equation}
\\
Eq(\ref{E5}) involves both quarks and anti-quarks. As in Ref\cite{Last} we use the same parametrization both for quarks and anti-quarks. Assuming the quark and anti-quark have equal normalization constants, we obtain for a specific flavor
\begin{equation}
\label{Eb}
e^{{D_0}}=\sum_{i=1}^{n_f}e^{2}_{i}\left(2 e^{D_0^i}\right)
\end{equation}
\\

From HERA data \cite{H1,ZE}, Eq(\ref{E6}) was fitted in Ref\cite{Last} with
\begin{eqnarray}
\label{E7}
D_0 &=& 0.339\pm 0.145 \nonumber \\
D_1 &=& 0.073\pm 0.001 \nonumber \\
D_2 &=& 1.013\pm 0.01 \nonumber \\
D_3 &=& -1.287\pm 0.01 \nonumber \\
Q_0^2 &=& 0.062\pm 0.01 \ {\text G\text e\text V^2}
\end{eqnarray}

in the kinematical region,
\begin{eqnarray}
\label{E8}
& & 6.2\times10^{-7}\leq x\leq 10^{-2} \nonumber \\
& & 0.045\leq Q^2 \leq 120 \ {\text G\text e\text V^2}
\end{eqnarray}
\\

However, the phenomenological analysis has one inherent limitation: due to the  negative value of $D_3$, Eq(\ref{E6}) develops a singularity at $x_0 \backsim 0.019$ \cite{DK4, DK5} as it satisfies the condition $1+D_3+D_1\log\frac{1}{x_0}=0$, contrary to the expectation of a physically viable form of structure function. 

We recently suggested self-similar models which are free from such singularity besides having $\log Q^2$ rise in structure function instead of power law rise in $Q^2$ as reported in Ref \cite{bs1,bsc}.\\


Below, therefore we outline the alternative way of constructing singularity free self-similar model.
explore alternative ways of making the model singularity free.

\subsection*{Model 2}
\subsubsection*{\textbf{An improved singularity free self-similarity based model of proton structure function at small \textit{x}}}
\label{mod1}
To get a singularity free self-similarity based model of proton structure function, one can redefine the magnification factor $M_1 = \left( 1+\dfrac{Q^2}{Q_0^2}\right)$ of Eq(\ref{E1}) as \cite{bs1, bsc} \\
\begin{equation}
\label{E300}
\hat{M}_1=\sum_{i=-n}^{n} \alpha_i M_1^i
\end{equation}
\\
reported in Ref \cite{bs1}. Only in a specific case, where $\alpha _1=1$ and all other coefficients cases vanish lead to the original $M_1$ as defined in Eq(\ref{E1}). If we take this generalization form of Eq(\ref{E300}) and if all the coefficients $\alpha _i (i=0, \ 1,\ 2, \ . \ . \ . \ , n)$ vanish then Eq(\ref{E300}) becomes

\begin{equation}
\label{E333}
\hat{M}_1=\sum_{j=1}^{n} \frac{B_j}{\left( 1+\frac{Q^2}{\hat{Q}_0^2}\right) ^j}
\end{equation}
\\
where
\begin{equation}
B_j= \alpha_{-j}
\end{equation}
\\
The defining uPDF therefore can be generalized to

\begin{equation}
\label{E311}
\log[M^2.\hat{f}_i(x,Q^2)]=\hat{D}_1\log\frac{1}{x}\log\hat{M}_1+\hat{D}_2\log\frac{1}{x}+\hat{D}_3\log\hat{M}_1+\hat{D}_0^i
\end{equation}
\\
instead of Eq(\ref{E1}), such that it will take the form

\begin{equation}
\label{E322}
\hat{f}_i(x,Q^2)=\frac{e^{\hat{D}_0^i}}{M^2}\left( \dfrac{1}{x}\right) ^{\hat{D}_2}\left( \hat{M}_1\right) ^{\hat{D}_3+\hat{D}_1\log\frac{1}{x}}
\end{equation}
\\
Taking only the two terms of Eq(\ref{E333}), $\hat{M_1}$ can be written as

\begin{equation}
\label{E344}
\hat{M}_1=\frac{B_1}{\left( 1+\frac{Q^2}{\hat{Q}_0^2}\right)}+\frac{B_2}{\left( 1+\frac{Q^2}{\hat{Q}_0^2}\right) ^2}
\end{equation}
\\
and the corresponding uPDF (Eq\ref{E322}) becomes

\begin{equation}
\label{E3m}
\hat{f}_i(x,Q^2)=\frac{e^{\hat{D}_0^i}}{M^2}\left( \dfrac{1}{x} \right)^{\hat{D}_2}\left( \frac{B_1}{\left( 1+\frac{Q^2}{\hat{Q}_0^2} \right)} \right)^{\hat{D}_3+\hat{D}_1\log\frac{1}{x}} \left( 1+ \frac{B_2}{B_1} \frac{1}{\left( 1+ \frac{Q^2}{\hat{Q}_0^2} \right) }  \right)^{\hat{D}_3+\hat{D}_1\log\frac{1}{x}}
\end{equation}
\\
Assuming the convergence of the polynomials as occurred in Eq(\ref{E3m}) and then integrate over $Q^2$, it yields the desired PDF

\begin{multline}
\label{E377}
\hat{q}_i(x,Q^2)=\frac{e^{\hat{D}_0^i} \hat{Q}_0^2}{M^2}\left( \frac{1}{x} \right)^{\hat{D}_2} \left( B_1 \right)^{\left( \hat{D}_3+\hat{D}_1\log\frac{1}{x} \right)}
\\
\left[  \frac{\left( \left(1+\frac{Q^2}{\hat{Q}_0^2} \right)^{\left( 1-\hat{D}_3-\hat{D}_1\log\frac{1}{x}\right) } -1 \right) }{\left( 1-\hat{D}_3-\hat{D}_1\log\frac{1}{x}\right) } -\frac{B_2}{B_1}\left(  \left(1+\frac{Q^2}{\hat{Q}_0^2} \right)^{\left( -\hat{D}_3-\hat{D}_1\log\frac{1}{x}\right) } -1 \right) \right] 
\end{multline}
\\
Using Eq(\ref{E377}) in Eq(\ref{E5}), the corresponding structure function, it gives 

\begin{multline}
\label{E388}
\hat{F}_2(x,Q^2)=\frac{e^{\hat{D}_0} \hat{Q}_0^2}{M^2}\left( \frac{1}{x} \right)^{\hat{D}_2-1} \left( B_1 \right)^{\left( \hat{D}_3+\hat{D}_1\log\frac{1}{x} \right)}
\\
\left[  \frac{\left( \left(1+\frac{Q^2}{\hat{Q}_0^2} \right)^{\left( 1-\hat{D}_3-\hat{D}_1\log\frac{1}{x}\right) } -1 \right) }{\left( 1-\hat{D}_3-\hat{D}_1\log\frac{1}{x}\right) } -\frac{B_2}{B_1}\left( \left(1+\frac{Q^2}{\hat{Q}_0^2} \right)^{\left( -\hat{D}_3-\hat{D}_1\log\frac{1}{x}\right) } -1 \right) \right] 
\end{multline}
\\
with the condition that 

\begin{equation}
\hat{D}_3+\hat{D}_1\log\frac{1}{x}\neq1
\end{equation}
as the equality will yield a undesired singularity. Further, if the model parameters $\hat{D}_1$ and $\hat{D}_3$ satisfy the additional condition

\begin{equation}
\label{1}
\hat{D}_3+\hat{D}_1\log\frac{1}{\hat{x}_0}=1
\end{equation}
then the resultant PDF

\begin{equation}
\label{E400}
\tilde{q}_i(x,Q^2)=\frac{e^{\tilde{D}_0^i} \tilde{Q}_0^2}{M^2}\left( \frac{1}{x} \right)^{\tilde{D}_2} \tilde{B}_1 \left[ \log \left( 1+\frac{Q^2}{\tilde{Q}_0^2} \right) -\frac{\tilde{B}_2}{\tilde{B}_1}\left(\frac{1}{\left( 1+\frac{Q^2}{\tilde{Q}_0^2}\right) }-1 \right)  \right] 
\end{equation}
\\
And the corresponding structure function is

\begin{equation}
\label{E411}
\tilde{F}_2(x,Q^2)=\frac{e^{\tilde{D}_0} \tilde{Q}_0^2}{M^2}\left( \frac{1}{x} \right)^{\tilde{D}_2-1} \tilde{B}_1 \left[ \log \left( 1+\frac{Q^2}{\tilde{Q}_0^2} \right) -\frac{\tilde{B}_2}{\tilde{B}_1}\left(\frac{1}{\left( 1+\frac{Q^2}{\tilde{Q}_0^2}\right) }-1 \right)  \right] 
\end{equation}
\\
which is completely free from singularity except for $\tilde{D}_2\geq1$ . Such singularity is, however, consistent with the usual Regge expectation \cite{reg,m,rg,coo,yu}. Besides it has also the logarithmic rise in $Q^2$. The model has now got 4 parameters: $\tilde{B}_1, \ \tilde{D}_2, \ \tilde{Q}_0^2, \ \tilde{D}_0^i$ which have been fitted by using the compiled HERA data \cite{HERA} and obtained the phenomenological range of validity of $Q^2$ and \textit{x} within \cite{bs1}
\begin{eqnarray}
& & 2\times10^{-5}\leq x\leq 0.4 \nonumber \\
& & 1.2\leq Q^2 \leq 800  \ {\text G\text e\text V^2}
\end{eqnarray}
which is quite large in comparative to earlier work of Ref\cite{Last}. The fitted parameters are given in Table \ref{T3}.

\begin{table}[!tbp]
\caption{\label{T3}%
\textit{Results of the fit of  $\tilde{F_2}$}}
\begin{ruledtabular}
\begin{tabular}{cccccc}
\textrm{$\tilde{D}_0$}&
\textrm{$\tilde{D}_2$}&
\textrm{$\tilde{B}_1$}&
\textrm{$\tilde{B}_2$}&
\textrm{$\tilde{Q}_0^2$(GeV$^2$)}&
\textrm{$\chi^2$/ndf} \\
\colrule
0.294\tiny${\pm 0.009}$ & 1.237\tiny${\pm 0.01}$ & 0.438\tiny${\pm 0.004}$ & 0.687\tiny${\pm 0.02}$ & 0.046\tiny${\pm 0.0004}$  & 0.60 \\
\end{tabular}
\end{ruledtabular}
\end{table}

\subsection*{Model 3}
\subsubsection*{\textbf{The large x extrapolated version of model 2: model 3}}
If the magnification factor $M_2 = \dfrac{1}{x}$ is also generalized to $\left(\dfrac{1}{x}-1\right)$ for large \textit{x} as suggested in Ref\cite{DK6} then one has defined uPDF as:

\begin{equation}
\label{E44}
\log[M^2.\bar{f}_i(x,Q^2)]= \bar{D}_1.\log\left( \frac{1}{x}-1\right).\log\left(1+\frac{Q^2}{\bar{Q}_0^2}\right)+\bar{D}_2.\log\left( \frac{1}{x}-1\right) +\bar{D}_3.\log\left(1+\frac{Q^2}{\bar{Q}_0^2}\right)+\bar{D}_0^i
\end{equation}
instead of Eq(\ref{E1}) which leads to

\begin{equation}
\bar{f}_i(x,Q^2)= \frac{e^{\bar{D}_0^i}}{M^2} \left( \frac{1}{x}-1\right)^{D_2}\left(1+\frac{Q^2}{\bar{Q}_0^2}\right)^{D_3+D_1\log\left( \frac{1}{x}-1\right)}
\end{equation}
Generalizing the magnification factor $\hat{M_1}$ as in Eq(\ref{E344}) and taking only the two terms and assuming the convergence of the polynomials occurring in the expression as in Eq(\ref{E3m}), we obtain the generalized uPDF as: 
\begin{multline}
\label{E45}
\bar{f}_i(x,Q^2)=\frac{e^{\bar{D}_0^i}}{M^2}\left( \frac{1}{x} \right)^{\bar{D}_2}\left(1-x \right)^{\bar{D}_2}\left( \frac{\bar{B}_1}{\left( 1+\frac{Q^2}{\bar{Q}_0^2} \right)} \right)^{\bar{D}_3+\bar{D}_1\log\frac{1}{x}+\bar{D}_1\log(1-x)} 
\\
\left( 1+ \frac{\bar{B}_2}{\bar{B}_1} \frac{\left( \bar{D}_3+\bar{D}_1\log\frac{1}{x}+\bar{D}_1\log(1-x) \right) }{\left( 1+ \frac{Q^2}{\bar{Q}_0^2} \right) }  \right)
\end{multline}
\\
And hence corresponding PDF$\left(\bar{q}_i\right)$  and structure function$\left( \bar{F}_2\right) $ will be

\begin{multline}
\label{E44a}
\bar{q}_i(x,Q^2)=\frac{e^{\bar{D}_0^i} \bar{Q}_0^2}{M^2}\left( \frac{1}{x} \right)^{\bar{D}_2}\left(1-x \right)^{\bar{D}_2}  \left( \bar{B}_1 \right)^{\left( \bar{D}_3+\bar{D}_1\log\frac{1}{x}+\bar{D}_1\log(1-x)\right)}
\\
\left[  \frac{\left( \left(1+\frac{Q^2}{\bar{Q}_0^2} \right)^{\left( 1-\bar{D}_3-\bar{D}_1\log\frac{1}{x}-\bar{D}_1\log(1-x)\right) } -1 \right) }{\left( 1-\bar{D}_3-\bar{D}_1\log\frac{1}{x}-\bar{D}_1\log(1-x)\right) } -\frac{\bar{B}_2}{\bar{B}_1}\left( \left(1+\frac{Q^2}{\bar{Q}_0^2} \right)^{\left( -\bar{D}_3-\bar{D}_1\log\frac{1}{x}-\bar{D}_1\log(1-x)\right) } -1 \right) \right] 
\end{multline}
and 
\begin{multline}
\label{E45a}
\bar{F}_2(x,Q^2)=\frac{e^{\bar{D}_0} \bar{Q}_0^2}{M^2}\left( \frac{1}{x} \right)^{\bar{D}_2-1} \left(1-x \right)^{\bar{D}_2-1}  \left( \bar{B}_1 \right)^{\left( \bar{D}_3+\bar{D}_1\log\frac{1}{x}+\bar{D}_1\log(1-x)\right)}
\\
\left[  \frac{\left( \left(1+\frac{Q^2}{\bar{Q}_0^2} \right)^{\left( 1-\bar{D}_3-\bar{D}_1\log\frac{1}{x}-\bar{D}_1\log(1-x)\right) } -1 \right) }{\left( 1-\bar{D}_3-\bar{D}_1\log\frac{1}{x}-\bar{D}_1\log(1-x)\right) } -\frac{\bar{B}_2}{\bar{B}_1}\left( \left(1+\frac{Q^2}{\bar{Q}_0^2} \right)^{\left( -\bar{D}_3-\bar{D}_1\log\frac{1}{x}-\bar{D}_1\log(1-x)\right) } -1 \right) \right] 
\end{multline}
Imposing the condition 
\begin{equation}
\label{2}
\bar{D}_3+\bar{D}_1\log\frac{1}{x}+\bar{D}_1\log(1-x)=1
\end{equation}
\\
will lead to corresponding UPDF, PDF and structure function as 

\begin{equation}
\label{E46}
\bar{f}_i'(x,Q^2)=\frac{e^{\bar{D}_0'^i}}{M^2}\left( \frac{1}{x} \right)^{\bar{D}'_2}\left(1-x \right)^{\bar{D}'_2}\left( \frac{\bar{B}'_1}{\left( 1+\frac{Q^2}{\bar{Q}_0'^2} \right)} \right) \left( 1+ \frac{\bar{B}'_2}{\bar{B}'_1} \frac{1}{\left( 1+ \frac{Q^2}{\bar{Q}_0'^2} \right) }  \right)
\end{equation}
Corresponding PDF

\begin{equation}
\label{E47}
\bar{q}'_i(x,Q^2)=\frac{e^{\bar{D}_0'^i} \bar{Q}_0'^2}{M^2}\left( \frac{1}{x} \right)^{\bar{D}'_2} \left(1-x \right)^{\bar{D}'_2} \bar{B}'_1 \left[ \log \left( 1+\frac{Q^2}{\bar{Q}_0'^2} \right) -\frac{\bar{B}'_2}{\bar{B}'_1}\left(\frac{1}{\left( 1+\frac{Q^2}{\bar{Q}_0'^2}\right) }-1 \right)  \right] 
\end{equation}
and corresponding structure function

\begin{equation}
\label{E48}
\bar{F}'_2(x,Q^2)=\frac{e^{\bar{D}'_0} \bar{Q}_0'^2}{M^2}\left( \frac{1}{x} \right)^{\bar{D}'_2-1}\left(1-x \right)^{\bar{D}'_2} \bar{B}'_1 \left[ \log \left( 1+\frac{Q^2}{\bar{Q}_0'^2} \right) -\frac{\bar{B}'_2}{\bar{B}'_1}\left(\frac{1}{\left( 1+\frac{Q^2}{\bar{Q}_0'^2}\right) }-1 \right)  \right] 
\end{equation}
which is our improved form and also has slower logarithmic raise in $Q^2$ with the large $x$ behavior 
\begin{equation}
\label{E43}
\lim_{x\to1} \bar{F}'_2(x,Q^2)=0
\end{equation}
consistent with QCD \cite{m,coo,yu,bro}.

For Model 3, the range of validity is obtained within: 
\begin{eqnarray}
& & 2\times10^{-5}\leq x\leq 0.4 \nonumber \\
& & 1.2\leq Q^2 \leq 1200  \ {\text G\text e\text V^2}
\end{eqnarray}
which is quite larger in comparative to earlier models 1 and 2. The fitted parameters for Model 3 are given in Table \ref{T4}. The number of data points of $\bar{F}'_2$ is 302.\\

\begin{table}[bp]
\caption{\label{T4}%
\textit{Results of the fit of $\bar{F}'_2$ model 3}}
\begin{ruledtabular}
\begin{tabular}{ccccccc}
\textrm{$\bar{D}'_0$}&
\textrm{$\bar{D}'_2$}&
\textrm{$\bar{B}'_1$}&
\textrm{$\bar{B}'_2$}&
\textrm{$\bar{Q}_0'^2$(GeV$^2$)}&
\textrm{$\chi^2$/ndf} \\
\colrule
0.335\tiny${\pm 0.003}$ & 1.194\tiny${\pm 0.0009}$ & 0.519\tiny${\pm 0.006}$ & 0.082\tiny${\pm 0.001}$ & 0.056\tiny${\pm 0.001}$ & 0.24 \\
\end{tabular}
\end{ruledtabular}
\end{table}

\subsection{TMDs in the self-similarity based models 1, 2 and 3}
\subsection*{Model 1}
In this subsection we outline the method of constructing TMDPDF from UPDF \cite{bs1}. The simplest way to introduce TMD in the self-similarity based models 1, 2 and 3 is by redefining the magnification factor in the $k_t^2$-space i.e. $\left(1+\dfrac{k_t^2}{k_0^2}\right)$ which can be written as:

\begin{equation}
\label{fi}
\log \left[  M^2 f_i(x,k_t^2)\right]  = D_1.\log\frac{1}{x}.\log\left(1+\frac{k_t^2}{k_0^2}\right)+D_2.\log\frac{1}{x}+D_3.\log\left(1+\frac{k_t^2}{k_0^2}\right)+D_0^i
\end{equation}
or 
\begin{equation}
\label{tmd}
f_i(x,k_t^2) = \frac{e^{D_0^i}}{M^2} \left( \frac{1}{x}\right) ^{D_1 \log\left(1+\frac{k_t^2}{k_0^2}\right)} \left( \frac{1}{x} \right)^{D_2} \left(1+\frac{k_t^2}{k_0^2}\right)^{D_3} 
\end{equation}
\\
to be compared with Eq(\ref{E1}) in $Q^2$-space. Here, $k_t^2$ is the square of the intrinsic transverse momentum of the parton which has corresponding \textit{x} as the longitudinal fraction. The parameters $D_1,\ D_2,\ D_3$ are same as determined from Deep Inelastic HERA structure function data as earlier. Redefining the PDF of Eq(\ref{E2}) in terms of $k_t^2$-space as:
\begin{equation}
\label{sx}
q_i(x,Q^2) = \int_0^{|k_t|^2<Q^2}dk_t^2 \ f_i(x,k_t^2)
\end{equation}
\\
with the cut off $|k_t|^2<Q^2$, one can obtain the expressions for integrated PDF and structure function. Thus this minimal extension of the approach to transverse structure of Proton keeps the results of the previous forms of parton distribution and structure function unchanged. 

Clearly, this can be done only in a specific model frame as noted in Refs. \cite{z1,z2,z3,ZA}. But it could be of interest to explore this approach to study $k_t$ dependence TMD $f_i(x,k_t^2)$ in the specific \textit{x} region where the approach the parameters have been fitted by using DIS data.
However, Eq(\ref{sx}) has deep theoretical limitation at the level of quantum field theory as noted by Collins \cite{27a}. 

Further, it has been found in recent years that the DIS experiment is not sufficient to obtain full transverse structure of the nucleon. Additional information is obtained from Semi Inclusive DIS (SIDIS) \cite{ZA} where one observes a hadron in the final stage. 
Such process is described by a fragmentation function $D_i(z_h,P_{ht};Q^2)$, which is analogous to the uPDF $f_i(x,k_t;Q^2)$ discussed earlier. Here, $z_h$ and $P_{ht}$ are the longitudinal momentum fraction and transverse momentum of the final hadron\textit{ h} with respect to the fragmenting parton. The present approach, however, can not be accommodated the fragmentation function.

\subsection*{Model 2}

Following the similar procedure , TMD corresponding to Eq(\ref{E3m}) can be written as
\begin{equation}
\label{E366}
\hat{f}_i(x,k_t^2)=\frac{e^{\hat{D}_0^i}}{M^2}\left( \frac{1}{x} \right)^{\hat{D}_2}\left( \frac{B_1}{\left( 1+\frac{k_t^2}{\hat{k}_{t0}^2} \right)} \right)^{\hat{D}_3+\hat{D}_1\log\frac{1}{x}} \left( 1+ \frac{B_2}{B_1} \frac{\left( \hat{D}_3+\hat{D}_1\log\frac{1}{x} \right) }{\left( 1+ \frac{k_t^2}{\hat{k}_{t0}^2} \right) }  \right)
\end{equation}
\\
If the parameters $\hat{D}_3$ and $\hat{D}_1$ satisfy the additional condition at

\begin{equation}
\label{11}
\hat{D}_3+\hat{D}_1\log\frac{1}{\hat{x}_0}=1
\end{equation}
then the resultant TMD will be 

\begin{equation}
\label{E399} 
\tilde{f}_i(x,k_t^2)=\frac{e^{\tilde{D}_0^i}}{M^2}\left( \frac{1}{x} \right)^{\tilde{D}_2}\left( \frac{\tilde{B}_1}{\left( 1+\frac{k_t^2}{\tilde{k}_{t0}^2} \right)} \right) \left( 1+ \frac{\tilde{B}_2}{\tilde{B}_1} \frac{1}{\left( 1+ \frac{k_t^2}{\tilde{k}_{t0}^2} \right) }  \right)
\end{equation}



\subsection*{Model 3}
Similarly the TMDs corresponding to the model 3 (Eq \ref{E45}) becomes

\begin{multline}
\label{E45xx}
\bar{f}_i(x,k_t^2)=\frac{e^{\bar{D}_0^i}}{M^2}\left( \frac{1}{x} \right)^{\bar{D}_2}\left(1-x \right)^{\bar{D}_2}\left( \frac{\bar{B}_1}{\left( 1+\frac{k_t^2}{\bar{Q}_0^2} \right)} \right)^{\bar{D}_3+\bar{D}_1\log\frac{1}{x}+\bar{D}_1\log(1-x)} 
\\
\left( 1+ \frac{\bar{B}_2}{\bar{B}_1} \frac{\left( \bar{D}_3+\bar{D}_1\log\frac{1}{x}+\bar{D}_1\log(1-x) \right) }{\left( 1+ \frac{k_t^2}{\bar{Q}_0^2} \right) }  \right)
\end{multline}

Imposing the condition 
\begin{equation}
\label{2x}
\bar{D}_3+\bar{D}_1\log\frac{1}{x}+\bar{D}_1\log(1-x)=1
\end{equation}
the resultant TMD will be 

\begin{equation}
\label{E46x}
\bar{f}_i'(x,k_t^2)=\frac{e^{\bar{D}_0'^i}}{M^2}\left( \frac{1}{x} \right)^{\bar{D}'_2}\left(1-x \right)^{\bar{D}'_2}\left( \frac{\bar{B}'_1}{\left( 1+\frac{k_t^2}{\bar{Q}_0'^2} \right)} \right) \left( 1+ \frac{\bar{B}'_2}{\bar{B}'_1} \frac{1}{\left( 1+ \frac{k_t^2}{\bar{Q}_0'^2} \right) }  \right)
\end{equation}

\subsection{Froissart bound compatible self-similarity based Proton structure functions with three magnification factors and power law rise in $Q^2$}
In this subsection, we outline the changes in TMDs if Froissart compatibility is also additionally imposed in the structure function.
\subsection*{Model 1$^{'}$}
In order to accommodate Froissart Bound in models of structure function based on self-similarity , three magnification factors are needed instead of two:
\begin{eqnarray}
\label{m3}
M_1 &=& \left(1+\frac{Q^2}{Q_0^2}\right) \nonumber \\
M_2 &=& \dfrac{1}{x} \nonumber \\
M_3 &=& \log \frac{1}{x}
\end{eqnarray}
In Ref\cite{dkc}, it was pointed out that if the scale factor $\log \frac{1}{x}$ is taking as the magnification factor in \textit{x}-space instead of $\frac{1}{x}$ then one can obtain the self-similar models of structure function in compatibility with Froissart $\log^2\frac{1}{x}$ behavior. However, in our more recent communication \cite{bsd}, we have shown that the conclusion of inference \cite{dkc} is only for the pdf due to the additional multiplicative factor \textit{x} in structure function. Instead, one needs two magnification factor $M_2$ and $M_3$ as defined in Eq(\ref{m3}). We therefore construct the updf , pdf and structure function as follows: \\ \\
uPDF\\
It is defined by
\begin{multline}
\log[M^2.\grave{f}_i(x,Q^2)]=\grave{D}_1 \log M_1 \log M_2 \log M_3 + \grave{D}_2 \log M_1 \log M_2 + \grave{D}_3 \log M_2 \log M_3 
\\
+ \grave{D}_4 \log M_1 \log M_3 + \grave{D}_5 \log M_1 + \grave{D}_6 \log M_2 + \grave{D}_7 \log M_3 + \grave{D}_0{^i}
\end{multline}
instead of Eq(\ref{E1}) leading to

\begin{multline}
\label{xc1}
\grave{f}_i(x,Q^2)= \dfrac{ e^{\grave{D}_0^i}}{M^2} \ \left(\dfrac{1}{x} \right) ^{\grave{D}_2 \log\left(1+\frac{Q^2}{\grave{Q}_0^2}\right)+\grave{D}_6} 
\\
\times \left(\log \dfrac{1}{x} \right) ^{\grave{D}_1 \log\left(1+\frac{Q^2}{\grave{Q}_0^2}\right) \log 1/x + \grave{D}_3 \log 1/x + \grave{D}_4\log\left(1+\frac{Q^2}{\grave{Q}_0^2}\right) + \grave{D}_7 }\left(1+\frac{Q^2}{\grave{Q}_0^2}\right)^{\grave{D}_5}
\end{multline}
\\
and the corresponding PDF can be written by using Eq(\ref{sx})

\begin{multline}
\label{x1}
\grave{q}_i(x,Q^2)= \frac{e^{\grave{D}_0^i}\ \grave{Q}_0^2\ (1/x)^{\grave{D}_6}\ \left( \log \frac{1}{x}\right) ^{\grave{D}_3 \log \frac{1}{x} + \grave{D}_7}}{M^2\left(1+\grave{D}_5+\grave{D}_2\log \frac{1}{x} +(\grave{D}_4+\grave{D}_1\log \frac{1}{x})\log \log \frac{1}{x}\right)} 
\\ \\
\times \left( (1/x)^{\grave{D}_2 \log\left(1+\frac{Q^2}{\grave{Q}_0^2}\right)} (\log 1/x)^{\log\left(1+\frac{Q^2}{\grave{Q}_0^2}\right)\left(\grave{D}_4+\grave{D}_1 \log \frac{1}{x}\right) }\left(1+\frac{Q^2}{\grave{Q}_0^2}\right)^{\grave{D}_5+1} -1\right) 
\end{multline}
For very small \textit{x} and large $Q^2$ , the second term of Eq(\ref{x1}) can be neglected, leading to

\begin{multline}
\label{x2}
\grave{q}_i(x,Q^2)= \frac{e^{\grave{D}_0^i}\ \grave{Q}_0^2\ (1/x)^{\grave{D}_2 \log\left(1+\frac{Q^2}{\grave{Q}_0^2}\right)+ \grave{D}_6}}{M^2\left(1+\grave{D}_5+\grave{D}_2\log \frac{1}{x} +(\grave{D}_4+\grave{D}_1\log \frac{1}{x})\log \log \frac{1}{x}\right)}
\\ \\
\times\left( \log \frac{1}{x}\right)^{\grave{D}_7 + \grave{D}_3 \log \frac{1}{x} + \left( \grave{D}_4+\grave{D}_1 \log \frac{1}{x}\right)\times \log\left(1+\frac{Q^2}{\grave{Q}_0^2}\right)} \ \left(1+\frac{Q^2}{\grave{Q}_0^2}\right)^{\grave{D}_5+1} 
\end{multline}
\\
from which one can define structure function as:

\begin{multline}
\label{x3}
\grave{F}_2(x,Q^2)= \frac{e^{\grave{D}_0}\ \grave{Q}_0^2\ (1/x)^{\grave{D}_2 \log\left(1+\frac{Q^2}{\grave{Q}_0^2}\right)+ \grave{D}_6-1}}{M^2\left(1+\grave{D}_5+\grave{D}_2\log \frac{1}{x} +(\grave{D}_4+\grave{D}_1\log \frac{1}{x})\log \log \frac{1}{x}\right)} 
\\ \\
\times\left( \log \frac{1}{x}\right) ^{\grave{D}_7 + \grave{D}_3 \log \frac{1}{x} + \left( \grave{D}_4+\grave{D}_1 \log \frac{1}{x}\right)\times \log\left(1+\frac{Q^2}{\grave{Q}_0^2}\right)} \ \left(1+\frac{Q^2}{\grave{Q}_0^2}\right)^{\grave{D}_5+1} 
\end{multline}
\\
which has total 9 parameters: $\grave{Q}_0^2$ and $\grave{D}_i$s with $i=$ 0 to 7. \\ \\
Eq(\ref{x3}) shows the proper Froissart saturation behavior in the structure function which is possible under the following conditions on the model parameters:

\begin{eqnarray}
\label{lab}
& (1) & \ \grave{D}_2 \log\left(1+\frac{Q^2}{\grave{Q}_0^2}\right)+ \grave{D}_6 = 1  \nonumber \\ 
& (2) & \ \grave{D}_7 + \grave{D}_3 \log \frac{1}{x} + \left( \grave{D}_4+\grave{D}_1 \log \frac{1}{x}\right)\times \log\left(1+\frac{Q^2}{\grave{Q}_0^2}\right) = 2 
\end{eqnarray}
Further if $\grave{D}_7, \ \grave{D}_3, \ \grave{D}_1\ll \grave{D}_4$ in Eq(\ref{lab}), then $\grave{D}_4=\dfrac{2-\grave{D}_7}{\log\left(1+\frac{Q^2}{\grave{Q}_0^2}\right)}$, the Froissart compatible structure function will be

\begin{equation}
\label{x4}
\grave{F}_2(x,Q^2)= \frac{e^{\grave{D}_0}\ \grave{Q}_0^2\ \left( \log \frac{1}{x}\right) ^2 \ \left(1+\frac{Q^2}{\grave{Q}_0^2}\right)^{\grave{D}_5+1}}{M^2\left(1+\grave{D}_5+\grave{D}_2\log \frac{1}{x} +(\grave{D}_4+\grave{D}_1\log \frac{1}{x})\log \log \frac{1}{x}\right)} 
\end{equation}
\\
which reduces the number parameters by 3. So Eq(\ref{x4}) results in a self-similarity based model of structure function compatible with Froissart bound having a power law growth in $Q^2$.\\

Using HERAPDF1.0 \cite{HERA}, Eq(\ref{x4}) is fitted as in Ref \cite{bsd} and found its phenomenological ranges of validity: $1.3\times 10^{-4}\leq x \leq 0.02$ and $6.5 \leq Q^2 \leq 90$ GeV$^2$ with the fitted parameters listed in Table \ref{c7t1}. 

\begin{table}[!bp]
\caption{\label{c7t1}%
\textit{Results of the fit of} $\grave{F}_2$, Eq(\ref{x4})}
\begin{ruledtabular}
\begin{tabular}{ccccccc}
\textrm{$\grave{D}_0$}&
\textrm{$\grave{D}_1$}&
\textrm{$\grave{D}_2$}&
\textrm{$\grave{D}_4$}&
\textrm{$\grave{D}_5$}&
\textrm{$Q_0''^2$(GeV$^2$)}&
\textrm{$\chi^2$/ndf} \\
\colrule
0.1006\tiny${\pm 0.003}$ & 0.028\tiny${\pm 0.0008}$ & -0.036\tiny${\pm 0.0001}$ & 3.585\tiny${\pm 0.05}$ & -0.857\tiny${\pm 0.01}$ & 0.060\tiny${\pm 0.001}$ & 0.11 \\
\end{tabular}
\end{ruledtabular}
\end{table}

\subsection*{Model 2$^{'}$}
The above observation of necessity of having 3 magnification factors can be applied to improved self-similarity based model 2 as well. In this case, we can construct another new set of magnification factors using the generalized magnification factor $\hat{M}_1$ Eq(\ref{E344}) along with $M_2$ and $M_3$.\\ \\
The defining equation of uPDF is now:
\begin{multline}
\log[M^2.\ddot{f}_i(x,Q^2)]= \ddot{D}_1 \log \hat{M}_1 \log M_2 \log M_3 + \ddot{D}_2 \log \hat{M}_1 \log M_2 + \ddot{D}_3 \log M_2 \log M_3 
\\
+ \ddot{D}_4 \log \hat{M}_1 \log M_3 + \ddot{D}_5 \log \hat{M}_1 + \ddot{D}_6 \log M_2 + \ddot{D}_7 \log M_3 + \ddot{D}_0{^i}
\end{multline}
instead of Eq(\ref{E1}) leading to

\begin{equation}
\label{vz1}
\ddot{f}_i(x,Q^2)= e^{\ddot{D}_0^i}\ \ddot{Q}_0^2\ \left(\dfrac{1}{x} \right) ^{\ddot{D}_6} \left( \log \frac{1}{x} \right) ^{\ddot{D}_3 \log \frac{1}{x}+\ddot{D}_7} \ddot{B}_1 \left[ \frac{1}{\left( 1+ \frac{Q^2}{\ddot{Q}_0^2}\right)} +\frac{\ddot{B}_2}{\ddot{B}_1} \frac{1}{\left( 1+ \frac{Q^2}{\ddot{Q}_0^2}\right)^2}\right]  
\end{equation}
\\
The corresponding PDF and structure function will have the forms

\begin{equation}
\ddot{q}_i(x,Q^2)= e^{\ddot{D}_0^i}\ \ddot{Q}_0^2\ (1/x)^{\ddot{D}_6} \left( \log \frac{1}{x} \right) ^{\ddot{D}_3 \log \frac{1}{x}+\ddot{D}_7} \ddot{B}_1 \left[ \log \left( 1+ \frac{Q^2}{\ddot{Q}_0^2}\right) - \frac{\ddot{B}_2}{\ddot{B}_1} \left( \frac{1}{\left( 1+\frac{Q^2}{\ddot{Q}_0^2}\right)}-1\right) \right]  
\end{equation}
and

\begin{multline}
\label{ddot}
\ddot{F}_2(x,Q^2)= e^{\ddot{D}_0}\ \ddot{Q}_0^2\ (1/x)^{\ddot{D}_6-1} \left( \log \frac{1}{x} \right) ^{\ddot{D}_3 \log \frac{1}{x}+\ddot{D}_7} 
\\
\times \ddot{B}_1 \left[ \log \left( 1+ \frac{Q^2}{\ddot{Q}_0^2}\right) - \frac{\ddot{B}_2}{\ddot{B}_1} \left( \frac{1}{\left( 1+\frac{Q^2}{\ddot{Q}_0^2}\right)}-1\right) \right]
\end{multline}
\\
respectively. Putting the extra conditions on the model parameters as
\begin{eqnarray}
& (1) & \ \ddot{D}_6-1=0 \nonumber \\
& (2) & \ \ddot{D}_3 \log \frac{1}{x}+\ddot{D}_7=2
\end{eqnarray}
\\
will give the Froissart like behavior in structure function of Eq(\ref{ddot}) a new form :

\begin{equation}
\label{x5}
\ddot{F}_2(x,Q^2)= e^{\ddot{D}_0}\ \ddot{Q}_0^2\ \log ^2 \left( 1/x\right)  \ \ddot{B}_1 \left[ \log \left( 1+ \frac{Q^2}{\ddot{Q}_0^2}\right) - \frac{\ddot{B}_2}{\ddot{B}_1} \left( \frac{1}{\left( 1+\frac{Q^2}{\ddot{Q}_0^2}\right)}-1\right) \right]
\end{equation}
\\

The Froissart compatible version of the model (Eq \ref{x5}) has now $\ln^2 \frac{1}{x}$ rise instead of power law in $\frac{1}{x}$ without changing the $\log Q^2$ rise (Eq \ref{E411}). \\

Now using the HERAPDF1.0 \cite{HERA}, Eq(\ref{x5}) is fitted and obtained its phenomenological ranges of validity within:  $1.3\times 10^{-4}\leq x \leq 0.02$ and $6.5 \leq Q^2 \leq 60$ GeV$^2$ and also obtained the model parameters which are given in Table \ref{c7t2} \cite{bsd}. \\

\begin{table}[tbp]
\caption{\label{c7t2}%
\textit{Results of the fit of $\ddot{F}_2$, Eq(\ref{x5})}}
\begin{ruledtabular}
\begin{tabular}{ccccc}
\textrm{$\ddot{D}_0$}&
\textrm{$\ddot{B}_1$}&
\textrm{$\ddot{B}_2$}&
\textrm{$\ddot{Q}_0^2$(GeV$^2$)}&
\textrm{$\chi^2$/ndf} \\
\colrule
0.00047\tiny${\pm 0.0003}$ & 0.056\tiny${\pm 0.002}$ & 0.672\tiny${\pm 0.02}$ & 0.022\tiny${\pm 0.001}$  & 0.27 \\
\end{tabular}
\end{ruledtabular}
\end{table}

\subsection*{Model 3$^{'}$}
If the third magnification factor $M_3$ is large-\textit{x} extrapolated: $\log \dfrac{1}{x} \rightarrow \log \left(\dfrac{1}{x} -1 \right)$ then the corresponding uPDF PDF and structure function becomes:
\\
uPDF

\begin{multline}
\label{vz3}
\breve{f}'_i(x,Q^2)= \frac{e^{\breve{D}_0'^i}}{M^2}\ (1/x)^{\breve{D}'_6}\ (1-x)^{\breve{D}'_6} \left( \log \frac{1-x}{x} \right) ^{\breve{D}'_3 \log \left( \frac{1}{x}-1\right) +\breve{D}'_7} 
\\
\times \breve{B}'_1 \left[ \frac{1}{\left( 1+ \frac{Q^2}{\breve{Q}_{0}'^2}\right)} +\frac{\breve{B}'_2}{\breve{B}'_1} \frac{1}{\left( 1+ \frac{Q^2}{\breve{Q}_{0}'^2}\right)^2}\right]  
\end{multline}
Corresponding PDF

\begin{multline}
\breve{q}'_i(x,Q^2)=e^{\breve{D}_0'^i}\ \breve{Q}_0'^2\ (1/x)^{\breve{D}'_6} (1-x)^{\breve{D}'_6} \left( \log \frac{1-x}{x} \right) ^{\breve{D}'_3 \log \left( \frac{1}{x}-1\right) +\breve{D}'_7} 
\\
\times \breve{B}'_1 \left[ \log \left( 1+ \frac{Q^2}{\breve{Q}_0'^2}\right)- \frac{\breve{B}'_2}{\breve{B}'_1} \left( \dfrac{1}{\left( 1+ \frac{Q^2}{\breve{Q}_0'^2}\right)}-1\right) \right]  
\end{multline}
and the structure function

\begin{multline}
\breve{F}'_2(x,Q^2)=e^{\breve{D}'_0}\ \breve{Q}_0'^2\ (1/x)^{\breve{D}'_6-1} (1-x)^{\breve{D}'_6} \left( \log \frac{1-x}{x} \right) ^{\breve{D}'_3 \log \left( \frac{1}{x}-1\right) +\breve{D}'_7} 
\\
\times \breve{B}'_1 \left[ \log \left( 1+ \frac{Q^2}{\breve{Q}_0'^2}\right) - \frac{\breve{B}'_2}{\breve{B}'_1} \left( \dfrac{1}{\left( 1+ \frac{Q^2}{\breve{Q}_0'^2}\right)}-1\right) \right]   
\end{multline}
\\
Putting the extra conditions
\begin{eqnarray}
& (1) & \ \breve{D}'_6-1=0 \nonumber \\
& (2) & \ \breve{D}'_3 \log \left( \frac{1}{x}-1\right)+\breve{D}'_7=2
\end{eqnarray}
\\
can show the Froissart like behavior in the structure function as: 
\\
\begin{equation}
\label{x7}
\breve{F}'_2(x,Q^2)= e^{\breve{D}'_0}\ \breve{Q}_0'^2\ (1-x)\ \log ^2 \left( \frac{1-x}{x}\right)  \ \breve{B}'_1 \left[ \log \left( 1+ \frac{Q^2}{\breve{Q}_0'^2}\right) - \frac{\breve{B}'_2}{\breve{B}'_1} \left( \dfrac{1}{\left( 1+ \frac{Q^2}{\breve{Q}_0'^2}\right)}-1\right) \right]  
\end{equation}
\\
to be compared with Eq(\ref{E48}) of the original model 3.\\

\begin{table}[!tbp]
\caption{\label{c7t4}%
\textit{Results of the fit of $\breve{F}'_2$, Eq(\ref{x7})}}
\begin{ruledtabular}
\begin{tabular}{ccccc}
\textrm{$\breve{D}'_0$}&
\textrm{$\breve{B}'_1$}&
\textrm{$\breve{B}'_2$}&
\textrm{$\breve{Q}_0'^2$(GeV$^2$)}&
\textrm{$\chi^2$/ndf} \\
\colrule
0.008\tiny${\pm 0.001}$ & 0.034\tiny${\pm 0.0008}$ & 0.251\tiny${\pm 0.01}$ & 0.057\tiny${\pm 0.005}$  & 0.26 \\
\end{tabular}
\end{ruledtabular}
\end{table}

Using the HERAPDF1.0 \cite{HERA}, Eq(\ref{x7}) is fitted and obtained its phenomenological ranges of validity within:  $1.3\times 10^{-4}\leq x \leq 0.02$ and $6.5 \leq Q^2 \leq 120$ GeV$^2$ with the obtained model parameters which are given in Table \ref{c7t4} \cite{bsd}.\\

\subsection{TMDs in Froissart compatible self-similarity based models 1$^{'}$, 2$^{'}$ and 3$^{'}$}
In this subsection, we outline the method of construction of TMDPDF from the corresponding UPDF of models 1$^{'}$, 2$^{'}$ and 3$^{'}$ (Eqs \ref{xc1}, \ref{vz1} and \ref{vz3})
\subsection*{Model 1$^{'}$}
Let us show TMD corresponding to Eq(\ref{xc1}) which has the form :

\begin{multline}
\label{cx2}
\grave{f}_i(x,k_t^2)= \dfrac{e^{\grave{D}_0^i}}{M^2} \ \left(\dfrac{1}{x} \right) ^{\grave{D}_2 \log\left(1+\frac{k_t^2}{\grave{k}_{0}^2}\right)+\grave{D}_6} 
\\
\times \left(\log \dfrac{1}{x} \right) ^{\grave{D}_1 \log\left(1+\frac{k_t^2}{\grave{k}_{0}^2}\right) \log \frac{1}{x} + \grave{D}_3 \log 1/x + \grave{D}_4\log\left(1+\frac{k_t^2}{\grave{k}_{0}^2}\right) + \grave{D}_7 }\left(1+\frac{k_t^2}{\grave{k}_{0}^2}\right)^{\grave{D}_5}
\end{multline}
Eq(\ref{cx2}) can show the proper Froissart bound like behavior in TMD under the following conditions:
\begin{eqnarray}
& (1) & \ \grave{D}_2 \log\left(1+\frac{k_t^2}{\grave{k}_{0}^2}\right)+ \grave{D}_6= 0  \nonumber \\ 
& (2) & \ \grave{D}_7 + \grave{D}_3 \log \frac{1}{x} + \left( \grave{D}_4+\grave{D}_1 \log \frac{1}{x}\right)\times \log\left(1+\frac{k_t^2}{\grave{k}_{0}^2}\right) = 2
\end{eqnarray}
\\
Further, if $\grave{D}_7, \ \grave{D}_3, \ \grave{D}_1\ll \grave{D}_4$, then $\grave{D}_4=\dfrac{2-\grave{D}_7}{\log\left(1+\frac{k_t^2}{\grave{k}_{0}^2}\right)}$, the Froissart Bound compatible TMD will be
\begin{equation}
\label{cx3}
\grave{f}_i(x,k_t^2)= \frac{e^{\grave{D}_0^i}}{M^2} \left(\log \dfrac{1}{x} \right)^{2} \left(1+\frac{k_t^2}{\grave{k}_{0}^2}\right)^{\grave{D}_5}
\end{equation}

\subsection*{Model 2$^{'}$}

Following the procedure as outlined earlier, the TMD corresponding to Eq(\ref{vz1}) in the present approach will be 

\begin{equation}
\label{cx5}
\ddot{f}_i(x,k_t^2)= \frac{e^{\ddot{D}_0^i}}{M^2}\ \left(\dfrac{1}{x} \right)^{\ddot{D}_6} \left( \log \frac{1}{x} \right) ^{\ddot{D}_3 \log \frac{1}{x}+\ddot{D}_7} \ddot{B}_1 \left[ \frac{1}{\left( 1+ \frac{k_t^2}{\ddot{k}_{0}^2}\right)} +\frac{\ddot{B}_2}{\ddot{B}_1} \frac{1}{\left( 1+ \frac{k_t^2}{\ddot{k}_{0}^2}\right)^2}\right]  
\end{equation}
Putting the extra conditions
\begin{eqnarray}
& (1) & \ \ddot{D}_6=0 \nonumber \\
& (2) & \ \ddot{D}_3 \log \frac{1}{x}+\ddot{D}_7=2
\end{eqnarray}
will give the Froissart like behavior in TMD as:
\begin{equation}
\label{cx6}
\ddot{f}_i(x,k_t^2)= \frac{e^{\ddot{D}_0^i}}{M^2}\ \left( \log \frac{1}{x} \right) ^{2} \ddot{B}_1 \left[ \frac{1}{\left( 1+ \frac{k_t^2}{\ddot{k}_{0}^2}\right)} +\frac{\ddot{B}_2}{\ddot{B}_1} \frac{1}{\left( 1+ \frac{k_t^2}{\ddot{k}_{0}^2}\right)^2}\right]  
\end{equation}

\subsection*{Model 3$^{'}$}
The TMD corresponding to Eq(\ref{vz3}) will be of the form :
\begin{multline}
\label{cx9}
\breve{f}'_i(x,k_t^2)= \frac{e^{\breve{D}_0'^i}}{M^2}\ \left(\dfrac{1}{x} \right)^{\breve{D}'_6}\ (1-x)^{\breve{D}'_6} \left( \log \frac{1-x}{x} \right) ^{\breve{D}'_3 \log \left( \frac{1}{x}-1\right) +\breve{D}'_7} 
\\
\times \breve{B}'_1 \left[ \frac{1}{\left( 1+ \frac{k_t^2}{k_{0}'^2}\right)} +\frac{\breve{B}'_2}{\breve{B}'_1} \frac{1}{\left( 1+ \frac{k_t^2}{\breve{k}_{0}'^2}\right)^2}\right]  
\end{multline}
\\
Putting the extra conditions
\begin{eqnarray}
& (1) & \ \breve{D}'_6=0 \nonumber \\
& (2) & \ \breve{D}'_3 \log \left( \frac{1}{x}-1\right)+\breve{D}'_7=2
\end{eqnarray}
can show the Froissart like behavior in TMD as:
\begin{equation}
\label{cx10}
\breve{f}'_i(x,k_t^2)= \frac{e^{\breve{D}_0'^i}}{M^2} \ \left( \log \frac{1-x}{x} \right) ^{2}  \breve{B}'_1 \left[ \frac{1}{\left( 1+ \frac{k_t^2}{\breve{k}_{0}'^2}\right)} +\frac{\breve{B}'_2}{\breve{B}'_1} \frac{1}{\left( 1+ \frac{k_t^2}{\breve{k}_{0}'^2}\right)^2}\right]  
\end{equation}
\\

Eqs(\ref{cx3}, \ref{cx6} and \ref{cx10}) are the main results of the present work. These equations have shown how the incorporation of Froissart bound in the structure function changes the behavior in TMDs.
\section{Results}
\label{res}
Let us now compare TMDs of models 1, 2 and 3 and $1^{'}$, $2^{'}$ and $3^{'}$ by showing its variance with $k_t^2$ and \textit{x}.

\subsection{Graphical representation of TMDs for models 1, 2 and 3}
To compare the TMDs for models 1, 2 and 3 (Eq \ref{tmd}, \ref{E399}, \ref{E46x}), we take the mean values of the model parameters from the respective tables for each model and calculate the $e^{D_0^i}$s. Below , we show the tables \ref{lk}, \ref{lkk} and \ref{lkl} with mean values of the parameters for models 1, 2 and 3 respectively  \\ \\
model 1: $e^{{D_0}^u}=1.008=e^{{D_0}^d}$ and $e^{{D_0}^s}=0.252=e^{{D_0}^c}$ \\
model 2: $e^{{\tilde{D}_0}^u}=0.964=e^{{\tilde{D}_0}^d}$ and $e^{{\tilde{D}_0}^s}=0.241=e^{{\tilde{D}_0}^c}$ \\
model 3: $e^{\bar{D}_0'^u}=1.004=e^{\bar{D}_0'^d}$ and $e^{\bar{D}_0'^s}=0.251=e^{\bar{D}_0'^c}$ \\

\begin{table}[!bp]
\caption{\label{lk}}%
\textit{Mean values taken from Eq \ref{E7} for model 1}
\begin{ruledtabular}
\begin{tabular}{ccccc}
\textrm{$D_0$}&
\textrm{$D_1$}&
\textrm{$D_2$}&
\textrm{$D_3$}&
\textrm{$k_0^2$ (GeV$^2$)} \\
\colrule
0.339 & 0.073 & 1.013 & -1.287 & 0.062 \\
\end{tabular}
\end{ruledtabular}
\end{table}

\begin{table}[!bp]
\caption{\label{lkk}}%
\textit{Mean values taken from Table \ref{T3} for model 2}
\begin{ruledtabular}
\begin{tabular}{ccccc}
\textrm{$\tilde{D}_0$}&
\textrm{$\tilde{D}_2$}&
\textrm{$\tilde{B}_1$}&
\textrm{$\tilde{B}_2$}&
\textrm{$\tilde{k}_0^2$ (GeV$^2$)}\\
\colrule
0.294 & 1.237 & 0.438 & 0.687 & 0.046 \\
\end{tabular}
\end{ruledtabular}
\end{table}

\begin{table}[!bp]
\caption{\label{lkl}}%
\textit{Mean values taken from Table \ref{T4} for model 3}
\begin{ruledtabular}
\begin{tabular}{ccccc}
\textrm{$\bar{D}'_0$}&
\textrm{$\bar{D}'_2$}&
\textrm{$\bar{B}'_1$}&
\textrm{$\bar{B}'_2$}&
\textrm{$\bar{k}_0'^2$ (GeV$^2$)}\\
\colrule
0.335 & 1.194 & 0.519 & 0.082 & 0.056 \\
\end{tabular}
\end{ruledtabular}
\end{table}

\begin{figure}[!tbp]
\subfloat[]{%
  \includegraphics[width=0.49\linewidth]{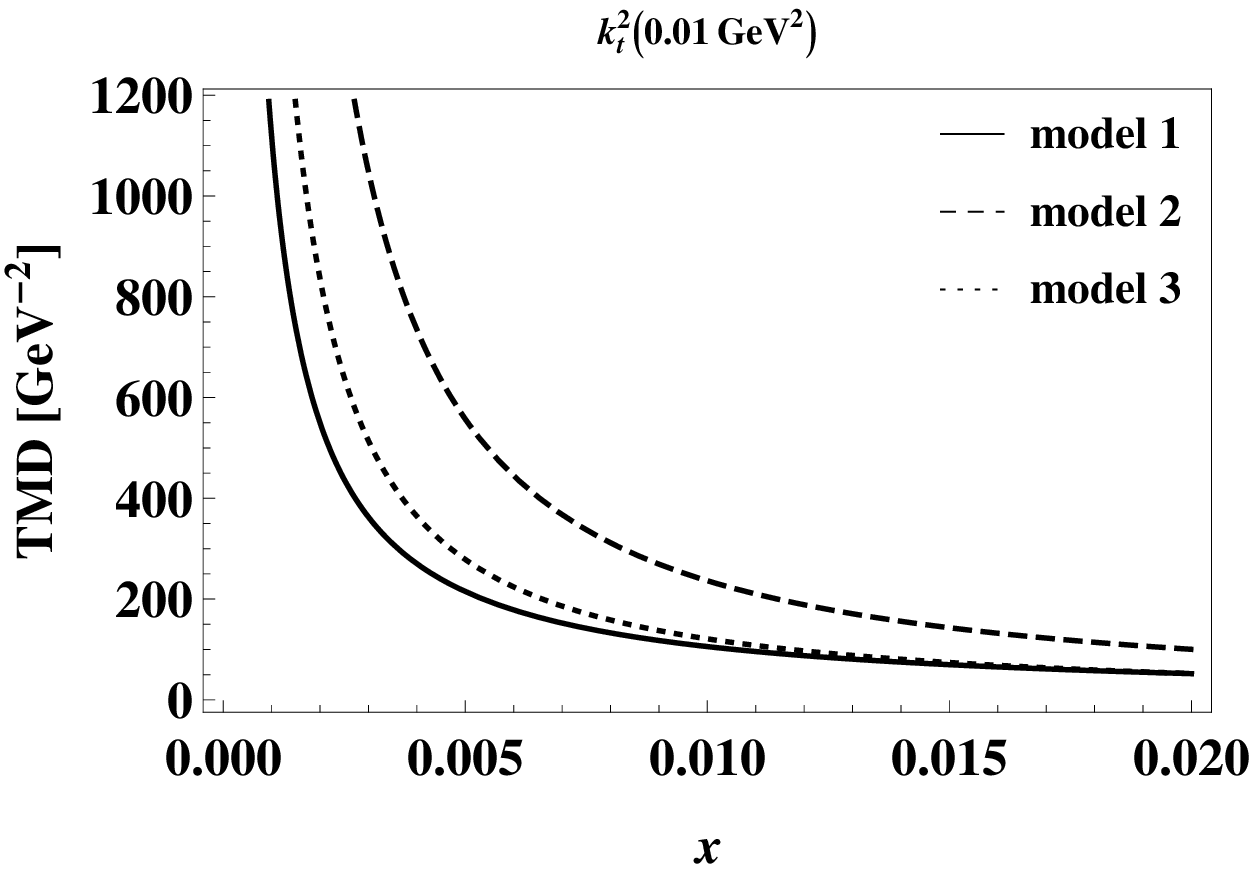}%
}\vspace{1ex}
\subfloat[]{%
  \includegraphics[width=0.48\linewidth]{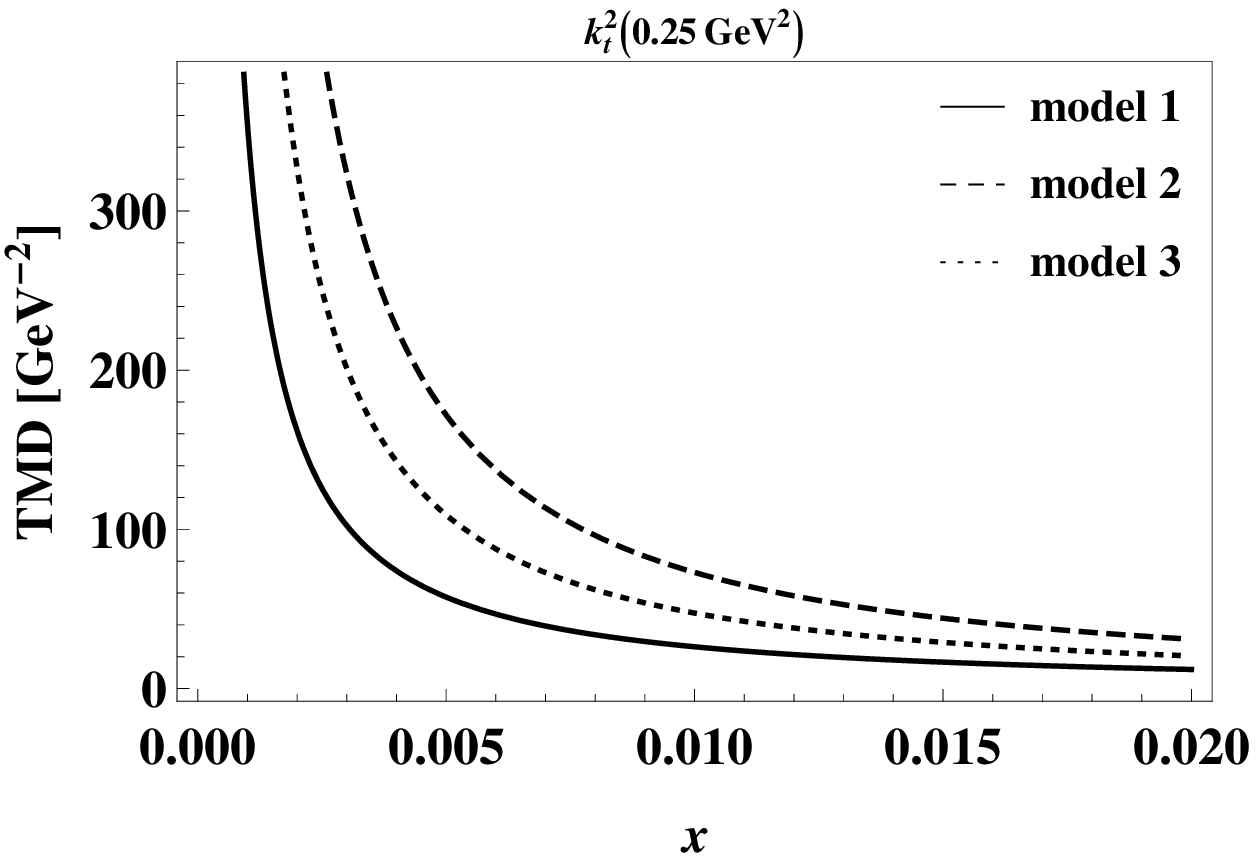}%
}\vspace{1ex}
\caption{TMD vs $x$ for two representative values of (a) $k_t^2=0.01$ GeV$^2$ and (b) $k_t^2=0.25$ GeV$^2$ for model 1, 2 and 3 taking only u and d quarks contributions.}
\label{y}
\end{figure}

\begin{figure}[bp]
\subfloat[]{%
  \includegraphics[width=0.49\linewidth]{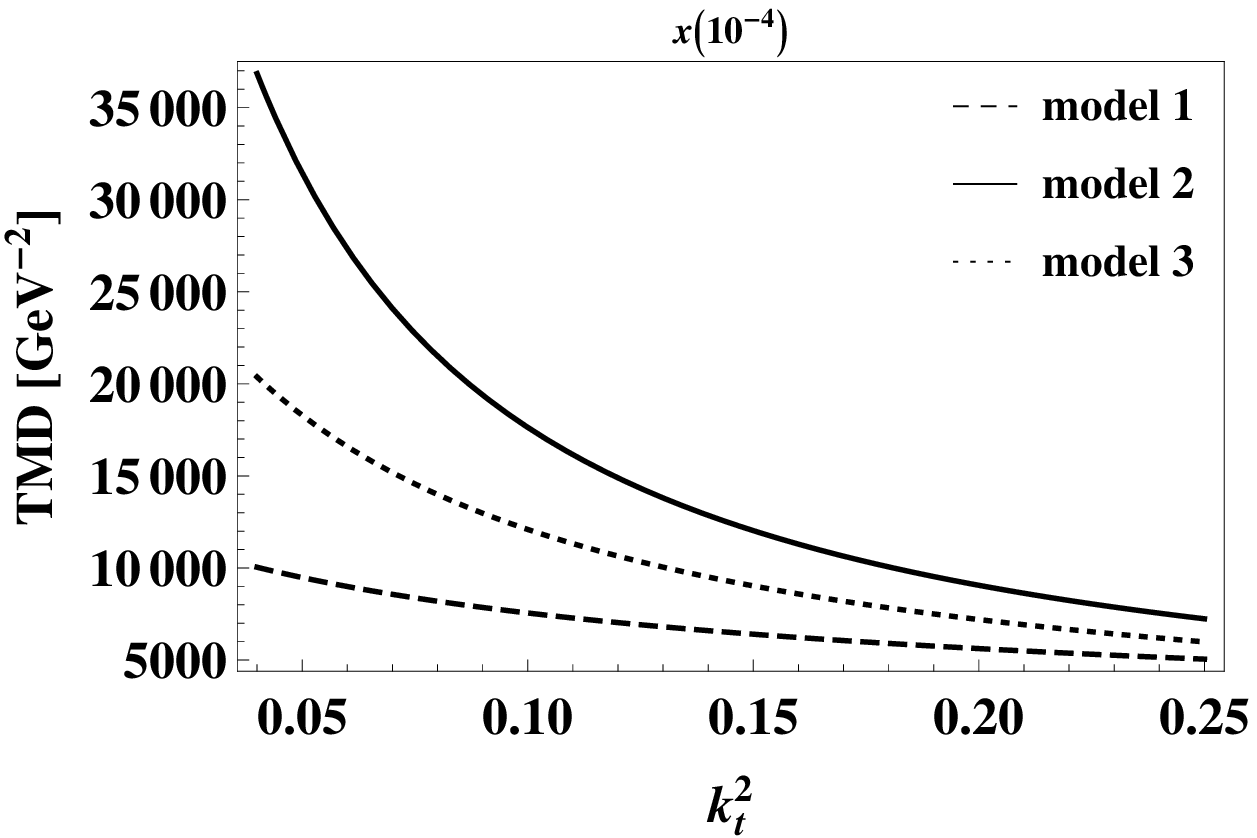}%
}\vspace{1ex}
\subfloat[]{%
  \includegraphics[width=0.48\linewidth]{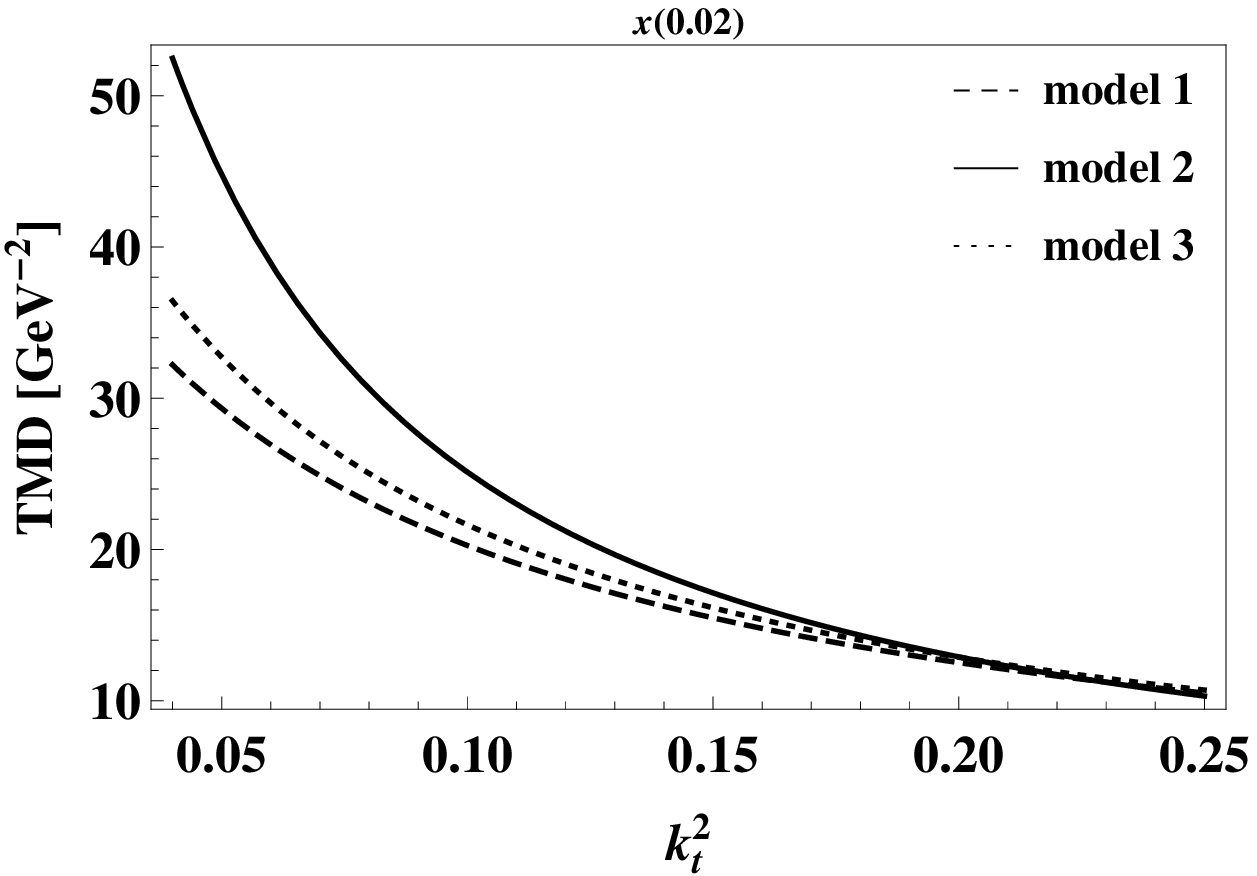}%
}\vspace{1ex}
\caption{TMD vs $k_t^2$ for two representative values of (a) $x=10^{-4}$ and (b) $x=0.02$ for model 1, 2 and 3 taking only u and d quarks contributions.}
\label{w}
\end{figure}

In Figs. \ref{y} and \ref{w}, we have shown TMDs vs \textit{x} and TMDs vs $k_t^2$ respectively for model 1 , 2 and 3. Graphical representation of TMDs are given within the ranges of \textit{x}: 10$^{-4}\leqslant x \leqslant 0.02$ and $k_t^2$: 0.01$\leqslant k_t^2 \leqslant 0.25$ GeV$^2$ for convenient.

From Fig. \ref{y} (a,b) we observe that for fixed $k_t^2$, as \textit{x} decreases TMD rises. Also the models 1, 2 and 3 have power law rise in $\frac{1}{x}$. The rise is faster for model 2 followed by model 1 and 3. The relative growth is determined by the magnitude of respective exponents of $\frac{1}{x}$ of Eqs (\ref{tmd}, \ref{E399}, \ref{E46x}).

Similarly, from Fig. \ref{w} (a, b) we observe that for fixed \textit{x}, the TMDs decreases as $k_t^2$ increases as expected for the same set of Eqs (\ref{tmd}, \ref{E399}, \ref{E46x}) showing polynomial fall with $\sim \frac{1}{k_t^2}$. The rate of fall determines the respective corresponding exponents of $\frac{1}{x}$ of Eqs (\ref{tmd}, \ref{E399}, \ref{E46x}).

\subsection{Graphical representation of TMDs for models $1^{'}$, $2^{'}$ and $3^{'}$}
To compare the Froissart saturated TMDs, we have calculated $e^{D_0^i}$s for each case: \\ \\
model $1^{'}$: \ \ \ $e^{{\grave{D}_0}^u}=0.904=e^{{\grave{D}_0}^d}$ and $e^{{\grave{D}_0}^s}=0.226=e^{{\grave{D}_0}^c}$ \\
model $2^{'}$: \ \ \ $e^{{\ddot{D}_0}^u}=0.818=e^{{\ddot{D}_0}^d}$ and $e^{{\ddot{D}_0}^s}=0.204=e^{{\ddot{D}_0}^c}$ \\
model $3^{'}$: \ \ \ $e^{{\breve{D}_0}'^u}=1.008=e^{{\breve{D}_0}'^d}$ and $e^{{\breve{D}_0}'^s}=0.252=e^{{\breve{D}_0}'^c}$ \\

and the mean values of the parameters for models $1^{'}$ $2^{'}$ $3^{'}$ are shown in Tables \ref{8t1}, \ref{8t2} and \ref{8t4} respectively.\\ \\
\begin{table}[!bp]
\caption{\label{8t1}}%
\textit{Mean values of the parameters of $\grave{f}_i$, Eq. \ref{cx3}; model $1^{'}$}
\begin{ruledtabular}
\begin{tabular}{cccccc}
\textrm{$\grave{D}_0$}&
\textrm{$\grave{D}_1$}&
\textrm{$\grave{D}_2$}&
\textrm{$\grave{D}_4$}&
\textrm{$\grave{D}_5$}&
\textrm{$\grave{k}_0^2$ (GeV$^2$)} \\
\colrule
0.1006 & 0.028 & -0.036 & 3.585 & -0.857 & 0.060   \\
\end{tabular}
\end{ruledtabular}
\end{table}

\begin{table}[!tbp]
\caption{\label{8t2}}%
\textit{Mean values of the parameters of $\ddot{f}_i$, Eq. \ref{cx6}; model $2^{'}$}
\begin{ruledtabular}
\begin{tabular}{cccc}
\textrm{$\ddot{D}_0$}&
\textrm{$\ddot{B}_1$}&
\textrm{$\ddot{B}_2$}&
\textrm{$\ddot{k}_0^2$ (GeV$^2$)} \\
\colrule
0.00047 & 0.056 & 0.672 & 0.022  \\
\end{tabular}
\end{ruledtabular}
\end{table}

\begin{table}[!tbp]
\caption{\label{8t4}}%
\textit{Mean values of the parameters of $\breve{f}'_i$, Eq. \ref{cx10}; model $3^{'}$}
\begin{ruledtabular}
\begin{tabular}{cccc}
\textrm{$\breve{D}'_0$}&
\textrm{$\breve{B}'_1$}&
\textrm{$\breve{B}'_2$}&
\textrm{$\breve{k}_0'^2$ (GeV$^2$)} \\
\colrule
0.008 & 0.034 & 0.251 & 0.057 \\
\end{tabular}
\end{ruledtabular}
\end{table}

\begin{figure}[!tbp]
\subfloat[]{%
  \includegraphics[width=0.49\linewidth]{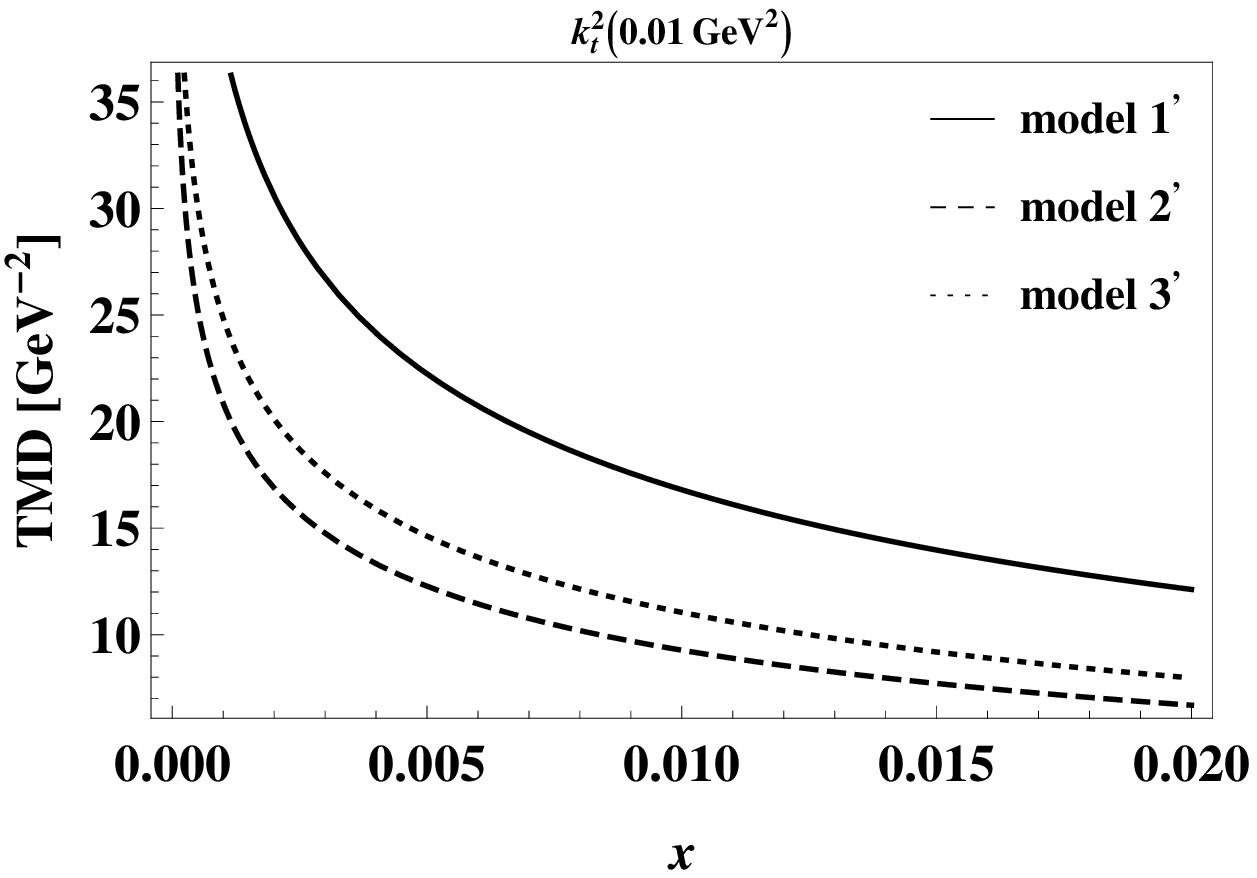}%
}\vspace{1ex}
\subfloat[]{%
  \includegraphics[width=0.48\linewidth]{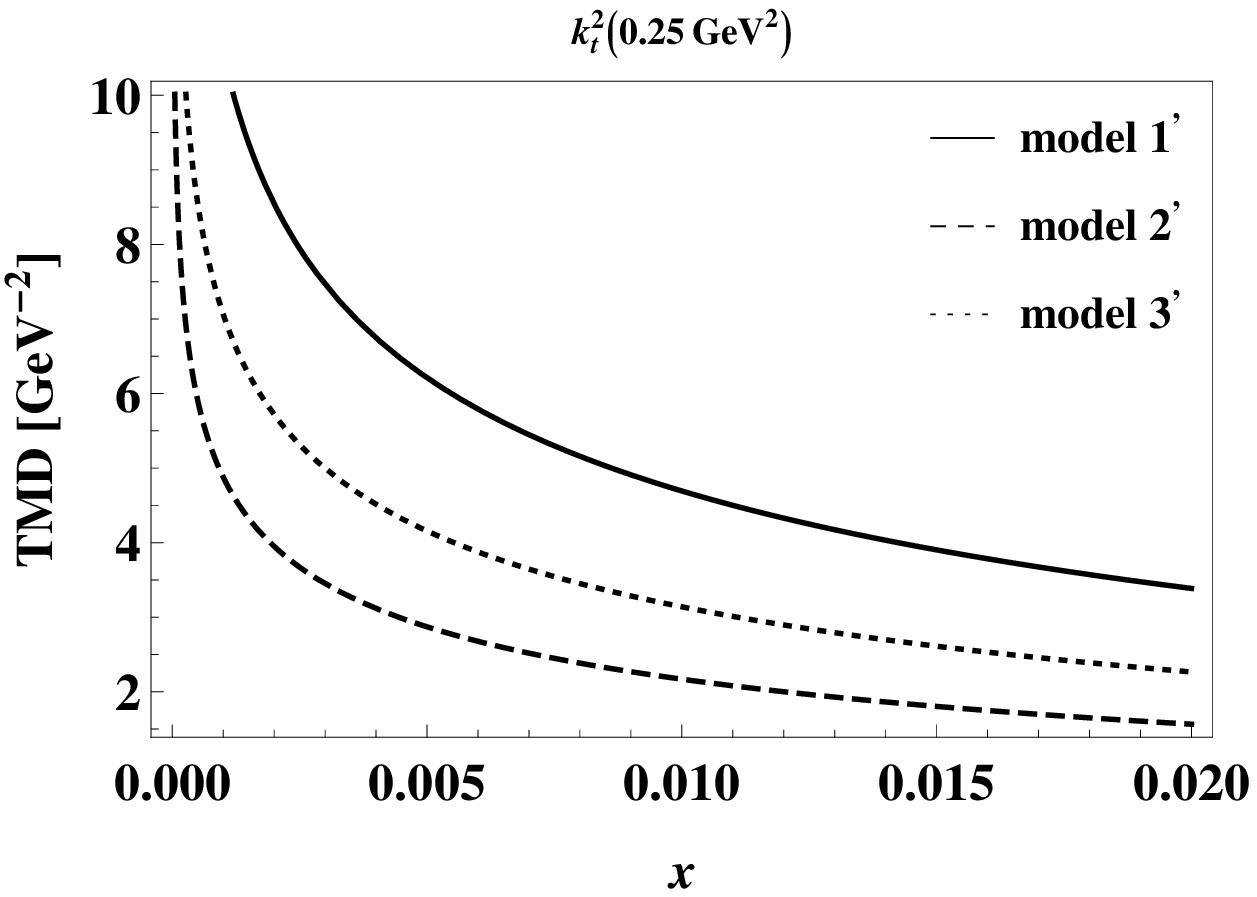}%
}\vspace{1ex}
\caption{TMD vs $x$ for two representative values of (a) $k_t^2=0.01$ GeV$^2$ and (b) $k_t^2=0.25$ GeV$^2$ for model $1^{'}$, $2^{'}$ and $3^{'}$ taking only u and d quarks contributions.}
\label{c8xx}
\end{figure}

\begin{figure}[!tbp]
\subfloat[]{%
  \includegraphics[width=0.49\linewidth]{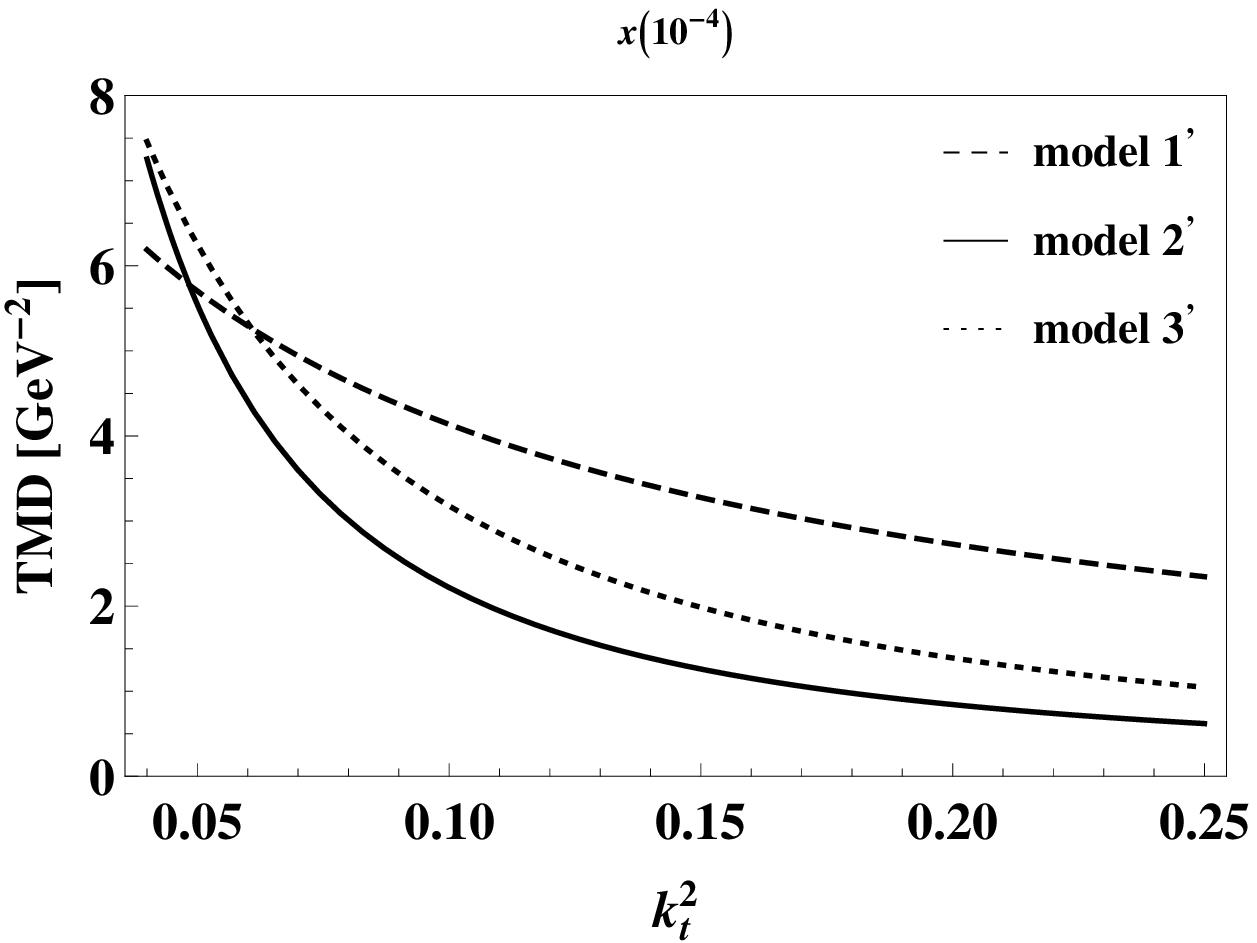}%
}\vspace{1ex}
\subfloat[]{%
  \includegraphics[width=0.5\linewidth]{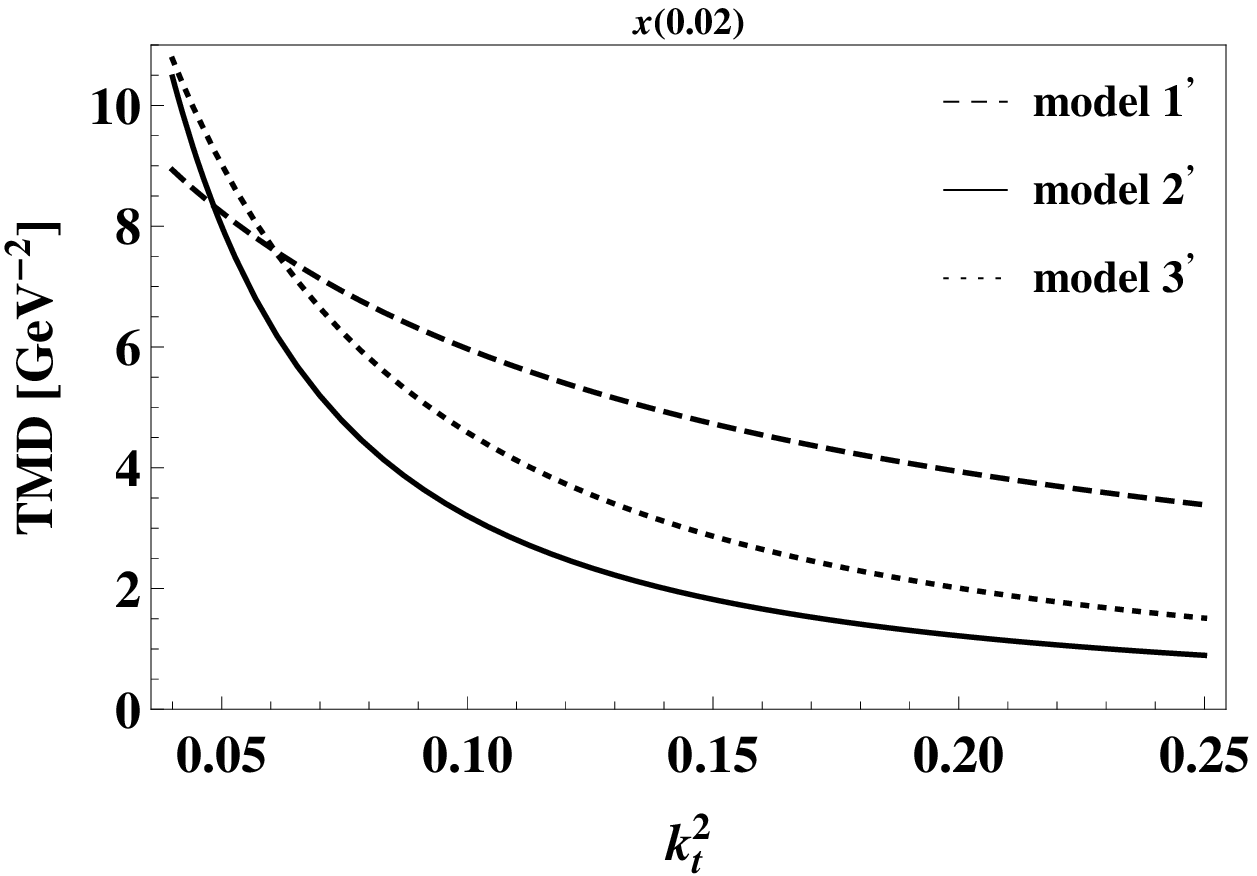}%
}\vspace{1ex}
\caption{TMD vs $k_t^2$ for two representative values of (a) $x=10^{-4}$ and (b) $x=0.02$ for model $1^{'}$, $2^{'}$ and $3^{'}$ taking only u and d quarks contributions.}
\label{c8fff}
\end{figure}

In Figs. \ref{c8xx} and \ref{c8fff}, we have shown TMDs vs \textit{x} and TMDs vs $k_t^2$ respectively for model $1^{'}$, $2^{'}$ and $3^{'}$. Graphical representation of TMDs are given within the ranges of \textit{x}: 10$^{-4}\leqslant x \leqslant 0.02$ and $k_t^2$: 0.01$\leqslant k_t^2 \leqslant 0.25$ GeV$^2$. From both the figures, we can observe from Fig \ref{c8xx} (a) and \ref{c8fff} (a), the expected pattern for fixed $k_t^2$, TMDs rises as \textit{x} decreases. But the rise is slower than the earlier Figs. \ref{y} and \ref{w} due to the softening of rise from $\frac{1}{x}$ to $\log ^2 \frac{1}{x}$. \\

From both the figures, we observe the expected behavior in \textit{x} for fixed $k_t^2$ (Fig. \ref{c8xx} a, b) and in $k_t^2$ for fixed \textit{x} (Fig. \ref{c8fff} a, b). For fixed $k_t^2$, rises in \textit{x} as $x\rightarrow 0$ is much slower than Fig. \ref{y} (a, b) due to slower $\log ^2 \frac{1}{x}$ rise to be compared with power law in $\frac{1}{x}$. 

The cross over points of Fig. \ref{c8fff} (a,b) at $k_t^2\approx 0.05$ GeV$^2$ is due to the new feature of models $1^{'}$, $2^{'}$ and $3^{'}$ where all the three models have identical value. This feature is absent in models 1, 2 and 3. \\

In several TMD models \cite{h2,h,h1}, \textit{x} and $k_t^2$ are parameters in factorisable form:
\begin{equation}
\label{E23}
f_i(x,k_t^2;Q^2)= q_i(x,Q^2) h(k_t^2)
\end{equation}
where $h(k_t^2)$ is the Gaussian of the form of
\begin{equation}
\label{E24}
h(k_t^2)= \frac{1}{\langle k_t^2\rangle} e^{-\frac{k_t^2}{\langle k_t^2\rangle}}
\end{equation}
with normalization constant
\begin{equation}
\label{E25}
\int h(k_t^2)dk_t^2= 1
\end{equation}

We note that while the model 1 does not have such factorization property, the models (2, 3, $1^{'}$, $2^{'}$ and $3^{'}$) satisfy this property. However, unlike Eq (\ref{E24}) they have power law fall in $k_t^2 \sim \frac{1}{k_t^{\alpha}}$ which is not Gaussian. In a sense, while the model 1 is closer to the models of Ref. \cite{z3,holo,33,34,35}, the rest of the models (2, $1^{'}$, $2^{'}$ and $3^{'}$)have similar to the ones in Ref. \cite{52,53,54,55,56,57,58,59,60,61}.

\section{Conclusion}
\label{conc}
In this work, we have discussed how one can introduce the Transverse Momentum Dependent Parton Distribution Function (TMD) in self-similarity based models of proton. We have obtained the proper Froissart bound condition in TMDs with the sets in three magnification factors. Graphical representation of TMDs with and without Froissart saturation has also been shown where one can observe that TMDs with the power law in $\dfrac{1}{x}$ (models 1, 2 and 3) rises faster at small \textit{x} than the the Froissart bound compatible TMDs with the power law in $\log\frac{1}{x}$ (models  $1^{'}$, $2^{'}$ and $3^{'}$).\\

Let us now discuss the limitations of the present approach:\\

First, the information of the TMDs can be obtained only in a limited range of \textit{x}, where the model parameters are fitted from the available DIS data.

Second, the Eq \ref{sx} relating the updf to TMD can at best be considered as an effective model ansatz in view of the analysis of Ref \cite{27a}.

The third one is it does incorporate fragmentation function and hence falls short off analysis fully SIDIS.

As the last limitation of the present approach, we note that in order to bring the self-similarity based models in compatibility with $\log^2 \frac{1}{x}$ behavior, we have to introduce two magnification factors $\frac{1}{x}$ and $\log \frac{1}{x}$, a feature beyond the notion of monofractality of the structure function in the space of \textit{x} and necessity of multifractality instead \cite{333}.




Finally, as noted in earlier publications Ref \cite{bs1, bsc} , self-similarity is not a general property of QCD and is not established properly, either theoretically or experimentally. In this work, we have merely made a use of fractal techniques to parametrize a multivarible function like structure function as a method of generalization as in Ref \cite{Last}. We have shown, under specific condition among the defining parameters, a slower logarithmic rise in $Q^2$ of structure function is achievable, which is closer to QCD expectation than the earlier power law growth of Ref \cite{Last} and has a wider phenomenological ranges of \textit{x }and $Q^2$. It implies, in a limited kinematical range, the notion of self-similarity makes some sense. However, unlike perturbative QCD where the corresponding Lagrangian is well defined, Feynmann rules are derivable and the asymptotic freedom can be established by using the Renormalization Group Equation leading to such $\log Q^2$ terms, it is beyond the scope of the present work and hence can not be considered as a first principle result. 

\section*{Acknowledgment}
One of the authors (B.S.) thanks the UGC-BSR for financial support.

\end{document}